\documentclass[aip,apl,reprint,amsmath,amssymb]{revtex4-1}

\usepackage{graphicx}
\usepackage{dcolumn}
\usepackage{bm}
\usepackage[utf8]{inputenc}
\usepackage[T1]{fontenc}
\usepackage{mathptmx}
\usepackage{xcolor}

\draft
\begin{document}

\title{Magnetic field-induced thermal emission tuning of InSb-based metamaterials in the terahertz frequency regime}

\author{Andrew Caratenuto}
\author{Yanpei Tian}
\affiliation{ 
Department of Mechanical and Industrial Engineering, Northeastern University, Boston, MA 02115, USA}
\author{Mauro Antezza}
\affiliation{ 
Laboratoire Charles Coulomb (L2C), UMR 5221 CNRS-Université de Montpellier, F-34095 Montpellier, France}
\affiliation{ 
Institut Universitaire de France, 1 rue Descartes, F-75231, Paris Cedex 05, France}
\author{Gang Xiao}
\affiliation{ 
Department of Physics, Brown University, Providence, RI 02912, USA}
\author{Yi Zheng}%
 \email{y.zheng@northeastern.edu.}
\affiliation{ 
Department of Mechanical and Industrial Engineering, Northeastern University, Boston, MA 02115, USA}
\affiliation{ 
Department of Electrical and Computer Engineering, Northeastern University, Boston, MA 02115, USA}

\date{\today}

\begin{abstract}
This work theoretically and analytically demonstrates the magnetic field-induced spectral radiative properties of photonic metamaterials incorporating both Indium Antimonide (InSb) and Tungsten (W) in the terahertz (THz) frequency regime. We have varied multiple factors of the nanostructures, including composite materials, layer thicknesses and surface grating fill factors, which impact the light-matter interactions and in turn modify the thermal emission of the metamaterials. We have proposed and validated a method for determining the spectral properties of InSb under an applied direct current (DC) magnetic field, and have employed this method to analyze how these properties can be dynamically tuned by modulating the magnitude of the field. For the first time, we have designed an InSb-W metamaterial exhibiting unity narrowband emission which can serve as an emitter for wavelengths around 55 $\mu$m (approximately 5.5 THz). Additionally, the narrowband emission of this metamaterial can be magnetically tuned in both bandwidth and peak wavelength with a \textcolor{black}{normal emissivity} close to unity.

\end{abstract}

\maketitle

Within the fields of photonics and electromagnetism, metamaterials have received increasing attention over traditional bulk materials\cite{Lin2014,Soukoulis2010}. Metamaterials are specially designed photonic material structures engineered to exhibit specific thermal and optical properties that are unique from those of naturally occurring materials\cite{Kildi2013}. A typical metamaterial is composed of extremely thin material layers, often modulated with micro- or nano-scale surface patterns comparable to the wavelengths of the incoming light. These features can modify the behaviors of the inter-surface phonon polaritons (SPhPs) or surface plasmon polaritons (SPPs) within a metamaterial. Because SPhPs and SPPs greatly impact the optical resonances of dielectrics and metals, respectively, the rich microstructures of the metamaterials yield diverse radiative responses in comparison to their bulk counterparts\cite{HillenbrandPhonons2002,HuberPhonon2005,BarnesPlasmon2003,CaldwellPlasmon2015,Lin2014,Pendry2006}.

Metamaterials can be designed to exhibit many different spectral behaviors in various regions in the electromagnetic spectra, and as such have many practical applications \cite{Oliveri2015,Holloway2012,Pendry2006,Kildi2013,Sarkar2019,Oliveri2015,Liu2015,Baqir2017,Dincer2014,Rufangura2016,Cai2007,Liu2007,Barbosa2015,Wang2015,Desmet2006,Bellisola2012,Montoya2017,Ogawa2012,Xu2020,Liu2010,Xu2017THz,Tonouchi2007,Son2019,Galoda2007,Zhong2019,Ghanekar2016APL,Ghanekar2018APL}. For example, metamaterials for the ultraviolet (UV) region have critical applications in biochemical sensing, defense, and communications\cite{Liu2015,Baqir2017}. In the visible region, applications include solar cells\cite{Dincer2014,Rufangura2016}, optical cloaking\cite{Cai2007,Liu2007}, and imaging\cite{Barbosa2015}. There are also extensive applications in the infrared (IR) region, particularly the near- and mid-IR, encompassing solar cells\cite{Wang2015}, medical treatments,\cite{Desmet2006} spectroscopy for medical research and cancer diagnoses\cite{Bellisola2012,Montoya2017}, and IR sensing applications\cite{Ogawa2012,Xu2020,Liu2010}.

Few metamaterial applications have materialized in the terahertz (THz) region, which covers 0.1 -- 10 THz (3000 -- 30 $\mu$m)\cite{Xu2017THz}. However, scientists have demonstrated the merit of the THz region in the areas of medical spectroscopy, security and communications\cite{Tonouchi2007,Son2019,Galoda2007,Zhong2019}. 

Besides simply creating metamaterials with desirable spectral properties for a given application, researchers have also had success in dynamically tuning the spectral properties of a given metamaterial, opening the door to further applications such as self-adaptive radiative cooling and thermophotovoltaics, as well as the potential for cost reduction\cite{Turpin2014,Cao2018,Qu2017}. This active tuning can be induced using many techniques, including via thermal\cite{Qu2017,Cao2018,Nguyen2020}, mechanical\cite{Xu2018-deformation,Wang2020,Su2020}, electrical\cite{Chen2020,Qi2020}, magnetic\cite{Moncada2015,Hu2015}, or optical\cite{Lee2020,Dicken2009} means. Some dynamic tuning techniques may be suitable for certain wavelength ranges and relatively ineffective in others\cite{Turpin2014,Li2019APL,Jia2016APL,Driscoll2008APL}. For example, the mechanical tuning of metamaterials has already been demonstrated in the visible\cite{Gutruf2016}, infrared\cite{Liu2020Nature} and THz\cite{Su2020} regions, while magnetic tuning has shown most promise in the THz region\cite{Moncada2015,Sharma2020,Li2013Thz}.

\begin{figure*}
\includegraphics[width=17cm]{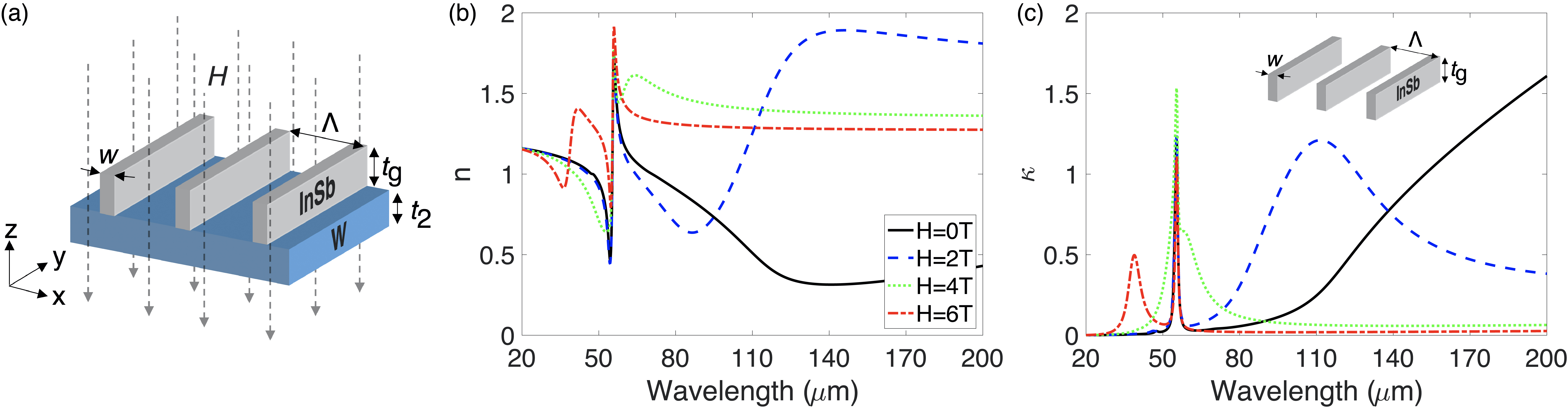}
\caption{\label{fig:Fig1} (a) Schematic of InSb grating on W thin film in the presence of an applied magnetic field. Magnetic field-induced in-plane refractive indices (b) $n$ and (a) $\kappa$ for InSb grating with $t_g = 7\mu$m, $\Lambda = 2\mu$m, and $\phi = 0.05$.}
\end{figure*}

This letter involves the magnetic tunability of metamaterials in the THz region. Here, an analytical expression for the dielectric function of the semiconductor material Indium Antimonide (InSb) is established. This expression takes into account the effect of an applied DC magnetic field, and as a result, the $n$ and $\kappa$ values of InSb are tuned when the magnetic field is present in the analysis. The introduction of the magnetic field also gives rise to interesting quantum energy transport\cite{Mauro1} and entanglement\cite{Mauro2} phenomena. Various InSb metamaterials are theoretically and analytically demonstrated using the established dielectric functions of thin films, multilayer structures, and surface gratings. Several InSb thicknesses and grating designs are studied to understand the impact of structure geometry on selective emission. The addition of a Tungsten (W) substrate to the metamaterials is also investigated with respect to the \textcolor{black}{normal emissivity} of each structure.

Notably, an InSb metamaterial with unity narrowband \textcolor{black}{emissivity} at 55 $\mu$m (approximately 5.5 THz) has been  demonstrated. Furthermore, the \textcolor{black}{emissivity} of this narrowband metamaterial is tuned via a magnetic field applied normal to the surface. The field modifies the peak wavelength and bandwidth of the metamaterial, and therefore allows both features to be actively tuned using an applied magnetic field.

We have analytically evaluated a series of metamaterial structures consisting of InSb and W, one of which is illustrated schematically in Fig.~\ref{fig:Fig1}a. This grating structure is defined by a grating layer thickness of $t_g$, grating width $w$, and grating period $\Lambda$, and is built upon a W film of thickness $t_2$. Structures are assumed to be infinite in the $x$- and $y$-directions. An applied magnetic field with a variable strength of $0-6$ T is oriented perpendicular to the W film (along the $z$-axis).

\textcolor{black}{For a generic reciprocal thermal emitter, emissivity can be expressed as
    $e_{\theta,\omega} = \frac{1}{2}\sum_{\mu=s,p}(1-|\tilde{R}_{\theta,\omega}^{(\mu)}|^2-|\tilde{T}_{\theta,\omega}^{(\mu)}|^2)$,
where $\theta$ is the incident angle and $\omega$ is the angular frequency of the emitted wave. The polarization-dependent effective Fresnel reflection and transmission coefficients are given by $\tilde{R}_{\theta,\omega}^{(\mu)}$ and $\tilde{T}_{\theta,\omega}^{(\mu)}$, respectively\cite{Ghanekar2016}. The superscript $\mu = s$ (or $p$) denotes the transverse electric (or magnetic) polarization of the incident waves. The reflection and transmission coefficients can be calculated using recursive Fresnel relations at each interface\cite{Tian2020}. For nonreciprocal emitters which have anisotropic permittivity tensors, such as those incorporating magneto-optical materials, Kirchhoff's Law is violated\cite{Zhang_Kirchoff,Zhu_Kirchoff,Zhao_Kirchoff}, and the absorptivity $\alpha_{\theta,\omega}$ cannot be assumed to be equal to the emissivity. In this case, the aforementioned emissivity equation is only valid for cases of normal incidence ($\theta = 0^\circ$). For this reason, this work focuses on normal emissivities of the selected metamaterials.}

\textcolor{black}{The normal $z$-component of the wave vector for medium $i$ is given by $k_{i,z}=\sqrt{\omega^2\epsilon_{i}(\omega)/c^2-k_{\rho}^2}$, where $\epsilon_{i}(\omega)$ is the relative permittivity of medium $i$ as a function of angular frequency $\omega$, $k_{\rho}$ is the magnitude of the in-plane wave vector ($k_{\rho}=0$ for normal incidence), and $c$ is the speed of light in vacuum\cite{Tian2020}.} The dielectric function is related to the real $(n)$ and imaginary $(\kappa)$ parts of the refractive index as $\sqrt{\epsilon}=n+j\kappa$, where $j$ is the imaginary unit\cite{Ghanekar2016}. The dielectric functions for the grating layers are determined using the second order effective medium theory, which is validated to hold true as the grating periods considered in this letter ($\leq 2$ $\mu$m) are significantly smaller than the wavelengths of interest ($\geq 20$ $\mu$m)\cite{Ghanekar2016}. Using this method, the effective transverse electric and transverse magnetic dielectric functions of the grating layers are given by\cite{Ghanekar2016}
\begin{equation}\label{effmed1}
    \epsilon_{TE,2}=\epsilon_{TE,0}\bigg[1+\frac{\pi^2}{3}\bigg(\frac{\Lambda}{\lambda}\bigg)^2\phi^2(1-\phi)^2\frac{(\epsilon_{A}-\epsilon_{B})^2}{\epsilon_{TE,0}}\bigg]
\end{equation}
and
\begin{eqnarray}\label{effmed2}
    \nonumber\epsilon_{TM,2}=\epsilon_{TM,0}\bigg[1+\frac{\pi^2}{3}\bigg(\frac{\Lambda}{\lambda}\bigg)^2\phi^2(1-\phi)^2(\epsilon_{A}\\
    -\epsilon_{B})^2\epsilon_{TE,0}\bigg(\frac{\epsilon_{TM,0}}{\epsilon_{A}\epsilon_{B}}\bigg)^2\bigg],
\end{eqnarray}
respectively, where $\Lambda$ is the grating period, $\lambda$ is the incident wavelength, and $\phi$ is the filling ratio of the grating defined as $w/\Lambda$. The relative permittivities of the two materials in the surface grating, InSb and air, are $\epsilon_{A}$ and $\epsilon_{B}$, respectively (with $\epsilon_{B}=1$). $\epsilon_{TE,0}$ and $\epsilon_{TM,0}$ are the zeroth order transverse electric and transverse magnetic effective dielectric functions, which are given by $\epsilon_{TE,0}=\phi\epsilon_{A}+(1-\phi)\epsilon_{B}$ and $\epsilon_{TM,0}=(\phi/\epsilon_{A}+(1-\phi)/{\epsilon_{B}})^{-1}$, respectively\cite{Ghanekar2016}.

In this work, the dielectric function of W is determined based on $n$ and $\kappa$ values from the literature\cite{Ordal1988}. In the absence of a magnetic field, the dielectric function for InSb is given by\cite{Palik1976}
\begin{equation}
    \epsilon=\epsilon_{\infty}\bigg(1+\frac{\omega_{L}^2-\omega_{T}^2}{\omega_{T}^2-\omega^2-j\Gamma\omega}-\frac{\omega_{p}^2}{\omega(\omega+j\gamma)}\bigg)
\end{equation}
where $\epsilon_{\infty}$ is the high-frequency dielectric constant, $\omega_{L}$ is the longitudinal optical phonon frequency, $\omega_{T}$ is the transverse optical phonon frequency, $\Gamma$ is the phonon damping constant, and $\gamma$ is the free-carrier damping constant. The plasma frequency $\omega_{p}$ of free carriers of density $N$ and effective mass $m^{*}$ is described by $\omega_{p}=Nq_{e}^2/m^{*}\epsilon_{0}\epsilon_{\infty}$, where $q_{e}$ and $\epsilon_{0}$ are the elementary charge and vacuum permittivity, respectively, and values for the InSb-specific constants are taken from the literature\cite{Moncada2015}.

Under a magnetic field $H$ applied along the $z$-direction, the permittivity tensor of InSb takes the following form\cite{Moncada2015}:
\begin{equation}\label{ematrix}
    \hat{\epsilon}(H)=\begin{pmatrix}
    \epsilon_{1}(H) & -j\epsilon_{2}(H) & 0\\[4pt]
    j\epsilon_{2}(H) & \epsilon_{1}(H) & 0\\[4pt]
    0 & 0 & \epsilon_{3}
    \end{pmatrix}
\end{equation}
where
\begin{subequations}
\begin{equation}\label{e1H}
    \epsilon_{1}(H)=\epsilon_{\infty}\bigg(1+\frac{\omega_{L}^2-\omega_{T}^2}{\omega_{T}^2-\omega^2-j\Gamma\omega}+\frac{\omega_{p}^2(\omega+j\gamma)}{\omega[\omega_{c}-(\omega+j\gamma)^2]}\bigg),
\end{equation}
\begin{equation}
    \epsilon_{2}(H)=\frac{\epsilon_{\infty}\omega_{p}^2\omega_c}{\omega[(\omega+j\gamma)^2-\omega_c^2]},
\end{equation}
\begin{equation}
    \epsilon_{3}=\epsilon_{\infty}\bigg(1+\frac{\omega_{L}^2-\omega_{T}^2}{\omega_{T}^2-\omega^2-j\Gamma\omega}-\frac{\omega_{p}^2}{\omega(\omega+j\gamma)}\bigg).
\end{equation}
\end{subequations}

The magnetic field contributes to the permittivity components $\epsilon_{1}(H)$ and $\epsilon_{2}(H)$ via the the cyclotron frequency term, which is given by $\omega_c=q_{e}H/m^{*}$. For the analyses herein, $q_{e}=1.602\times10^{-19}$ C, and $m^{*}=2.004\times10^{-32}$ kg. Using this method, we have calculated the wavelength-dependent $n$ and $\kappa$ values of $\epsilon_{1}(H)$ for an InSb grating structure ($t_g = 7\mu$m, $\Lambda = 2\mu$m, and $\phi = 0.05$) subjected to various magnetic field strengths, as shown in Fig.~\ref{fig:Fig1}b and Fig.~\ref{fig:Fig1}c, respectively. It can be seen that the magnetic field has a dramatic effect on these optical functions.

Finally, with the anisotropy of the permittivity tensor as expressed in Eq.~\ref{ematrix}, the equation for the normal $z$-component of the wave vector can be re-written as $k_{o,z}(H)=\sqrt{\omega^2\epsilon_{1}(H)/c^2-k_{\rho}^2}$ for ordinary waves and $k_{e,z}(H)=\sqrt{\omega^2\epsilon_{1}(H)/c^2-k_{\rho}^2\epsilon_{1}(H)/\epsilon_{3}}$
for extraordinary waves\cite{Moncada2015}. Combining this method with the aforementioned equations, we have calculated the \textcolor{black}{normal emissivity} for a variety of metamaterials incorporating InSb as both a grating layer and a thin film layer under an applied magnetic field.

\begin{figure}[ht]
    \includegraphics[width=8.5cm]{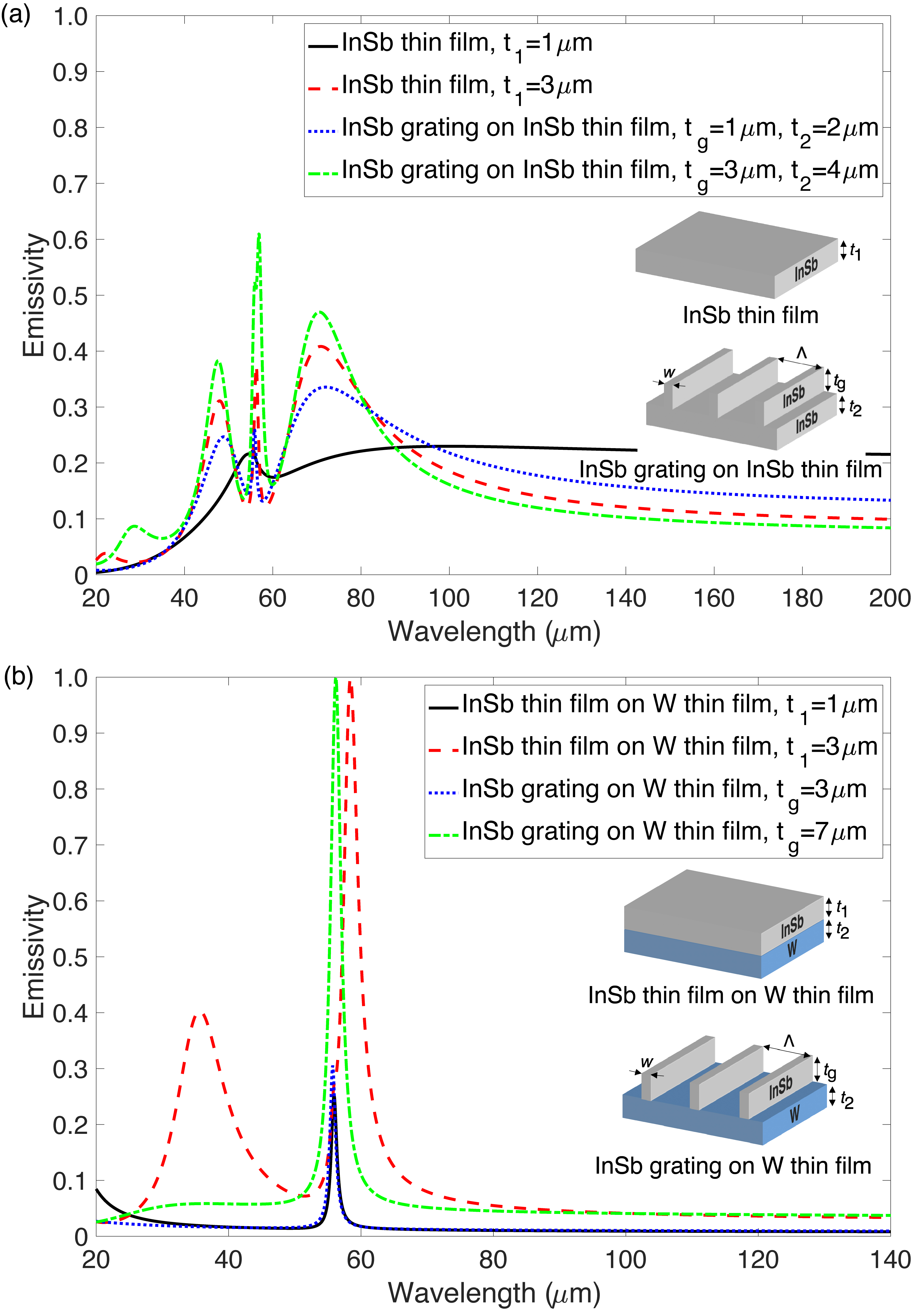}
    \caption{\textcolor{black}{Normal emissivities} of (a) InSb thin film and grating structures and (b) InSb-W thin film and grating structures, with $\Lambda=2{\mu}m$ and $\phi=0.05$. For (b), $t_2$ = 10 ${\mu}m$.}
    \label{fig:fig2ab}
\end{figure}

\begin{figure*}[ht]
    \includegraphics[width=17cm]{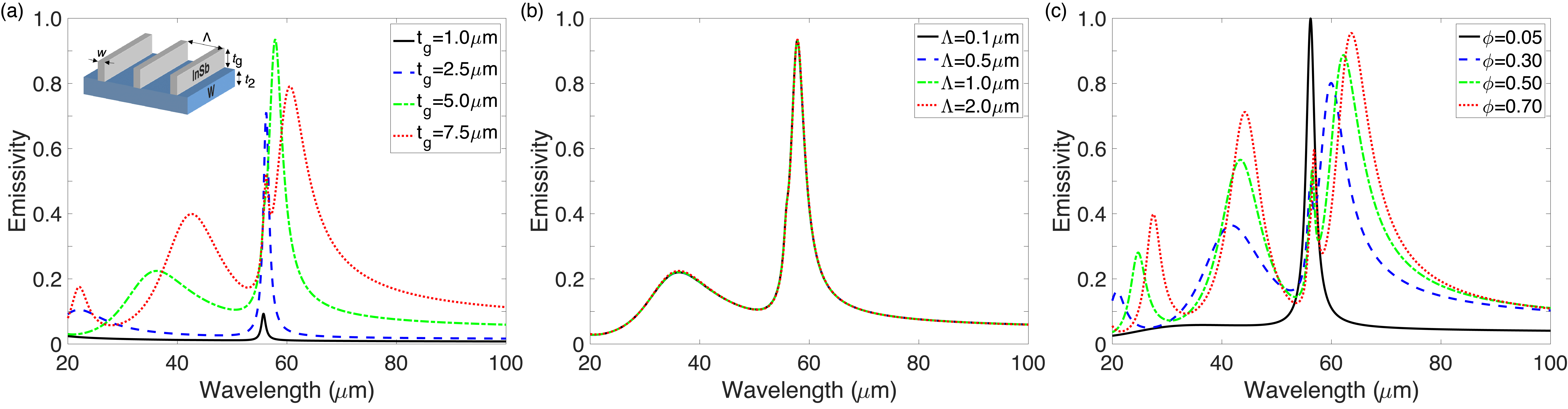}
    \caption{\textcolor{black}{Normal emissivities} of InSb-W grating structures with varying values of (a) grating thickness $t_g$, (b) grating period $\Lambda$, and (c) grating fill factor $\phi$. For (a), $\Lambda$ $=2$ $\mu$m and $\phi$ $=0.3$; for (b), $t_g$ $=5$ $\mu$m and $\phi$ $=0.3$; for (c), $t_g$ $=7$ $\mu$m and $\Lambda$ $=2$ $\mu$m. $t_2$ $=10$ $\mu$m for all cases.}
    \label{fig:gratingvary}
\end{figure*}

In the absence of the magnetic field, the \textcolor{black}{emissivity} as a function of wavelength is shown in Fig.~\ref{fig:fig2ab}a for selected InSb thin film and grating designs. An emission spike is present around 55 $\mu$m, which aligns with the major resonant wavelengths of $n$ and $\kappa$ in the zero-field case shown in Figs.~\ref{fig:Fig1}b and~\ref{fig:Fig1}c. The \textcolor{black}{emissivity} is observed to scale with thickness. However, neither changes in thickness nor the geometric features of the grating modify the SPhP resonances of the InSb structures in these cases. The pure InSb structures have obvious resonances in the $20-100$ $\mu$m wavelength region, yet their potential application as an emitter is hampered by relatively low emissivities and a lack of focus on a specific wavelength.

In contrast to the pure InSb metamaterials, we calculate the emissivities of InSb thin film and grating layers atop W base layers in Fig.~\ref{fig:fig2ab}b. The W base suppresses many of the smaller oscillations before and after the prominent 55 $\mu$m peak for the thin film cases, which we attribute to the introduction of SPPs from the W layer. The light-matter interactions are further modified when the InSb grating structures are analyzed in Fig.~\ref{fig:fig2ab}b. The chosen geometric factors suppress the pre-peak resonances, leaving only a pronounced narrowband at the major InSb resonance wavelength. With a grating thickness of $t_g=7$ $\mu$m, the unity narrowband emission of the InSb-W grating structure represents an exemplary design for a THz emitter around 55 $\mu$m.

The analyses in Fig.~\ref{fig:gratingvary} further detail the effects of grating parameters $t_g$, $\Lambda$ and $\phi$ on the \textcolor{black}{normal emissivity} of the InSb-W metamaterial. For larger values of $t_g$ in Fig.~\ref{fig:gratingvary}a, the InSb thickness increases the surface resonance wavelengths, as well as the \textcolor{black}{emissivity} in most cases. However, when $t_g$ is scaled to a specific level ($7.5$ $\mu$m in this case), the resonance specific to the $t_g$ alterations moves past the original InSb resonance, the latter of which is still apparent. This indicates that the resonances corresponding to modifications in $t_g$ and the original zero-field InSb resonance identified in Figs.~\ref{fig:Fig1}b and~\ref{fig:Fig1}c are not entirely interdependent and can be aligned to achieve strong emitter performance.

The \textcolor{black}{emissivity} of this metamaterial has a very weak dependence on the grating period $\Lambda$, as displayed in Fig.~\ref{fig:gratingvary}b. Modifying the period does not change the grating thickness, though it does alter the amount of InSb in a given unit area. Therefore, the fact that there is an extremely minor effect aligns with previous analyses as well as Eqs.~\ref{effmed1} and~\ref{effmed2}. For this reason, metamaterial design with InSb gratings can be performed with a large variety of periods without significant changes in performance, expanding the potential accessibility of this metamaterial.

Based on Eqs.~\ref{effmed1} and~\ref{effmed2}, the relative permittivity of a grating is heavily dependent upon the fill factor $\phi$. This is exemplified in Fig.~\ref{fig:gratingvary}c, where both resonant wavelength and \textcolor{black}{emissivity} are tuned by varying the fill factor. As the fill factor increases, the amount of InSb in the grating increases, while the amount of air in the grating is reduced. Hence, the gratings with reduced fill factors exhibit less apparent minor InSb resonances, although the zero-field InSb resonance remains prominent. This technique is applied to design the metamaterial exhibiting permanent narrowband \textcolor{black}{emissivity} in Fig.~\ref{fig:gratingvary}c ($\phi=0.05$), which has great potential for an InSb-based THz emitter. Larger fill factors may be useful for geometrically tuning peak location since the largest peak \textcolor{black}{emissivity} is not greatly reduced, but this approach adds many minor \textcolor{black}{emissivity} peaks which would reduce the efficiency of such an emitter. 

The three cases in Fig.~\ref{fig:gratingvary} show that the largest factor affecting the \textcolor{black}{emissivity} of the InSb-W grating structure is, generally, the amount of InSb present on top of the structure. When InSb is added, either by increasing the grating layer thickness or the fill factor, the peak magnitude and surface polariton resonance location tends to increase, and vice versa. This corresponds with the relatively small effect that varying the period has, since the ratio of InSb to air in a given unit cell is kept constant.

\begin{figure}[h!]
    \includegraphics [width=8.0cm]{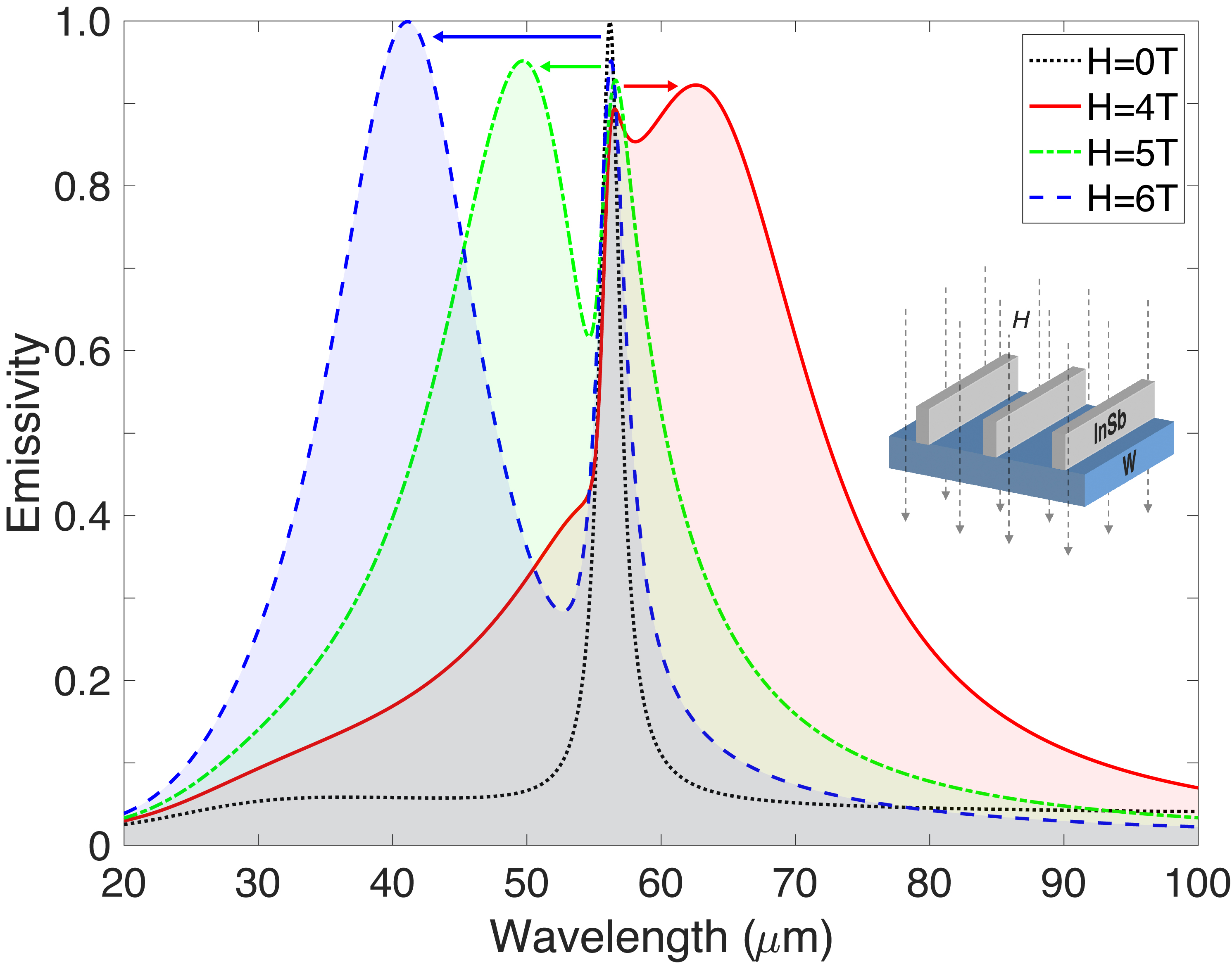}
    \caption{InSb-W grating structure, with $t_g=7$ ${\mu}m$, $\Lambda$ $=2$ $\mu$m, $\phi$ $=0.05$ and $t_2=10$ ${\mu}m$ under external magnetic fields of varying magnitudes. Field-induced curves are color-filled to highlight the broadband red or blue emissivity shifts brought on by the magnetic field. Arrows denote the direction of the dominant resonance shift from the original narrowband peak to longer (red) or shorter (blue) wavelengths.}
    \label{fig4}
\end{figure}

Finally, we have studied the dramatic effect of the magnetic field on the narrowband \textcolor{black}{emissivity} of an InSb-W grating structure. Figure~\ref{fig4} shows the \textcolor{black}{normal emissivity} of a grating structure with $t_g=7$ ${\mu}m$, $\Lambda$ $=2$ $\mu$m, and $\phi$ $=0.05$ under applied fields of 0, 4, 5 and 6 T. Tesla-scale magnetic fields significantly modify the light-matter interactions by widely broadening the resonant wavelengths over a range of 25 $\mu$m, while still maintaining the near-unity \textcolor{black}{emissivity} resonance peak. The \textcolor{black}{emissivity} initially red-shifts at lower magnetic field values, before being blue-shifted when the field intensity increases. As noted earlier, the applied magnetic field affects only the cyclotron frequency term $\omega_c$ in Eq.~\ref{e1H}, which enables it to modify the \textcolor{black}{emissivity} of the metamaterial. The peak shift at these field values is attributed to the cyclotron resonance line carrying the free-carrier dispersion to different frequencies, which occurs when $\omega_c$ becomes comparable to $\omega_T$\cite{Palik1976}. For this InSb-W metamaterial, the applied magnetic field broadens the bandwidth and shifts the resonant wavelength of the \textcolor{black}{emissivity}. At large values of $H$, the magnetic field modifies the values of $\epsilon_1(H)$ in Eq.~\ref{e1H} such that the extraordinary wave dispersion becomes hyperbolic\cite{Moncada2015}. This effect generates new broadband modes similar to those of SPPs which can be tuned by modifying the magnitude of $H$, but does not affect the original InSb narrowband resonance -- results which are consistent with previous studies\cite{Moncada2015,Palik1976}. These behaviors open new avenues for manipulating the optical emission properties of metamaterials, as well as remotely sensing magnetic fields in hard-to-reach spaces, such as inside high-energy particle detectors or electrical power systems.

In summary, we have investigated the optical emissive properties of InSb-based metamaterial structures, both with and without a W substrate, under the influence of a variety of geometric factors. We have also examined the same properties of InSb-W metamaterials under an applied magnetic field. A method for analytically expressing the dielectric function of InSb has been combined with the effective medium theory to enable this study. In the absence of the magnetic field, the volume fraction of the top InSb layer of the structure has the most prominent effect on the tunable \textcolor{black}{normal emissivity}. The inclusion of a W substrate suppresses many of the minor peaks while amplifying the most prominent resonance. This enhancement of the \textcolor{black}{emissivity} at a specific wavelength could be taken advantage of in many photonic applications. We attribute these behaviors to changes in surface polariton resonances arising from metamaterial structural variations. We have discovered that an applied magnetic field has the capability to tune the \textcolor{black}{emissivity} of an InSb-W grating structure, which has demonstrated unity narrowband \textcolor{black}{emissivity} around 55 $\mu$m in a zero field case. Through its effect on the cyclotron resonances, the external field is highly effective in red- and blue-shifting the peak \textcolor{black}{emissivity} wavelength and widening the narrowband to a broadband, while still preserving its near-unity magnitude. These findings may benefit applications in the THz region, as well as magnetic sensing in electrical systems, non-destructive sensing and imaging\cite{Galoda2007}, pharmaceuticals\cite{Ayala2007}, and astronomical observation\cite{Tonouchi2007,Betz1996}.

\begin{acknowledgments}
This project is supported by the National Science Foundation through grant number CBET-1941743.
\end{acknowledgments}

\section*{Availability of Data}
The data that support the findings of this study are available from the corresponding author upon reasonable request.

\nocite{*}
\section*{References}

\begin{thebibliography}{64}%
\makeatletter
\providecommand \@ifxundefined [1]{%
 \@ifx{#1\undefined}
}%
\providecommand \@ifnum [1]{%
 \ifnum #1\expandafter \@firstoftwo
 \else \expandafter \@secondoftwo
 \fi
}%
\providecommand \@ifx [1]{%
 \ifx #1\expandafter \@firstoftwo
 \else \expandafter \@secondoftwo
 \fi
}%
\providecommand \natexlab [1]{#1}%
\providecommand \enquote  [1]{``#1''}%
\providecommand \bibnamefont  [1]{#1}%
\providecommand \bibfnamefont [1]{#1}%
\providecommand \citenamefont [1]{#1}%
\providecommand \href@noop [0]{\@secondoftwo}%
\providecommand \href [0]{\begingroup \@sanitize@url \@href}%
\providecommand \@href[1]{\@@startlink{#1}\@@href}%
\providecommand \@@href[1]{\endgroup#1\@@endlink}%
\providecommand \@sanitize@url [0]{\catcode `\\12\catcode `\$12\catcode
  `\&12\catcode `\#12\catcode `\^12\catcode `\_12\catcode `\%12\relax}%
\providecommand \@@startlink[1]{}%
\providecommand \@@endlink[0]{}%
\providecommand \url  [0]{\begingroup\@sanitize@url \@url }%
\providecommand \@url [1]{\endgroup\@href {#1}{\urlprefix }}%
\providecommand \urlprefix  [0]{URL }%
\providecommand \Eprint [0]{\href }%
\providecommand \doibase [0]{http://dx.doi.org/}%
\providecommand \selectlanguage [0]{\@gobble}%
\providecommand \bibinfo  [0]{\@secondoftwo}%
\providecommand \bibfield  [0]{\@secondoftwo}%
\providecommand \translation [1]{[#1]}%
\providecommand \BibitemOpen [0]{}%
\providecommand \bibitemStop [0]{}%
\providecommand \bibitemNoStop [0]{.\EOS\space}%
\providecommand \EOS [0]{\spacefactor3000\relax}%
\providecommand \BibitemShut  [1]{\csname bibitem#1\endcsname}%
\let\auto@bib@innerbib\@empty
\bibitem [{\citenamefont {Lin}\ \emph {et~al.}(2014)\citenamefont {Lin},
  \citenamefont {Fan}, \citenamefont {Hasman},\ and\ \citenamefont
  {Brongersma}}]{Lin2014}%
  \BibitemOpen
  \bibfield  {author} {\bibinfo {author} {\bibfnamefont {D.}~\bibnamefont
  {Lin}}, \bibinfo {author} {\bibfnamefont {P.}~\bibnamefont {Fan}}, \bibinfo
  {author} {\bibfnamefont {E.}~\bibnamefont {Hasman}}, \ and\ \bibinfo {author}
  {\bibfnamefont {M.~L.}\ \bibnamefont {Brongersma}},\ }\href@noop {}
  {\bibfield  {journal} {\bibinfo  {journal} {Science}\ }\textbf {\bibinfo
  {volume} {345}},\ \bibinfo {pages} {298} (\bibinfo {year}
  {2014})}\BibitemShut {NoStop}%
\bibitem [{\citenamefont {Soukoulis}\ and\ \citenamefont
  {Wegener}(2010)}]{Soukoulis2010}%
  \BibitemOpen
  \bibfield  {author} {\bibinfo {author} {\bibfnamefont {C.~M.}\ \bibnamefont
  {Soukoulis}}\ and\ \bibinfo {author} {\bibfnamefont {M.}~\bibnamefont
  {Wegener}},\ }\href@noop {} {\bibfield  {journal} {\bibinfo  {journal}
  {Science}\ }\textbf {\bibinfo {volume} {330}},\ \bibinfo {pages} {1633}
  (\bibinfo {year} {2010})}\BibitemShut {NoStop}%
\bibitem [{\citenamefont {Kildishev}, \citenamefont {Boltasseva},\ and\
  \citenamefont {Shalaev}(2013)}]{Kildi2013}%
  \BibitemOpen
  \bibfield  {author} {\bibinfo {author} {\bibfnamefont {A.~V.}\ \bibnamefont
  {Kildishev}}, \bibinfo {author} {\bibfnamefont {A.}~\bibnamefont
  {Boltasseva}}, \ and\ \bibinfo {author} {\bibfnamefont {V.~M.}\ \bibnamefont
  {Shalaev}},\ }\href@noop {} {\bibfield  {journal} {\bibinfo  {journal}
  {Science}\ }\textbf {\bibinfo {volume} {339}} (\bibinfo {year}
  {2013})}\BibitemShut {NoStop}%
\bibitem [{\citenamefont {Hillenbrand}, \citenamefont {Taubner},\ and\
  \citenamefont {Keilmann}(2002)}]{HillenbrandPhonons2002}%
  \BibitemOpen
  \bibfield  {author} {\bibinfo {author} {\bibfnamefont {R.}~\bibnamefont
  {Hillenbrand}}, \bibinfo {author} {\bibfnamefont {T.}~\bibnamefont
  {Taubner}}, \ and\ \bibinfo {author} {\bibfnamefont {F.}~\bibnamefont
  {Keilmann}},\ }\href@noop {} {\bibfield  {journal} {\bibinfo  {journal}
  {Nature}\ }\textbf {\bibinfo {volume} {418}},\ \bibinfo {pages} {159}
  (\bibinfo {year} {2002})}\BibitemShut {NoStop}%
\bibitem [{\citenamefont {Huber}\ \emph {et~al.}(2005)\citenamefont {Huber},
  \citenamefont {Ocelic}, \citenamefont {Kazantsev},\ and\ \citenamefont
  {Hillenbrand}}]{HuberPhonon2005}%
  \BibitemOpen
  \bibfield  {author} {\bibinfo {author} {\bibfnamefont {A.}~\bibnamefont
  {Huber}}, \bibinfo {author} {\bibfnamefont {N.}~\bibnamefont {Ocelic}},
  \bibinfo {author} {\bibfnamefont {D.}~\bibnamefont {Kazantsev}}, \ and\
  \bibinfo {author} {\bibfnamefont {R.}~\bibnamefont {Hillenbrand}},\
  }\href@noop {} {\bibfield  {journal} {\bibinfo  {journal} {Applied Physics
  Letters}\ }\textbf {\bibinfo {volume} {87}} (\bibinfo {year}
  {2005})}\BibitemShut {NoStop}%
\bibitem [{\citenamefont {Barnes}, \citenamefont {Dereux},\ and\ \citenamefont
  {Ebbesen}(2003)}]{BarnesPlasmon2003}%
  \BibitemOpen
  \bibfield  {author} {\bibinfo {author} {\bibfnamefont {W.~L.}\ \bibnamefont
  {Barnes}}, \bibinfo {author} {\bibfnamefont {A.}~\bibnamefont {Dereux}}, \
  and\ \bibinfo {author} {\bibfnamefont {T.~W.}\ \bibnamefont {Ebbesen}},\
  }\href@noop {} {\bibfield  {journal} {\bibinfo  {journal} {Nature}\ }\textbf
  {\bibinfo {volume} {424}},\ \bibinfo {pages} {824} (\bibinfo {year}
  {2003})}\BibitemShut {NoStop}%
\bibitem [{\citenamefont {Caldwell}\ \emph {et~al.}(2015)\citenamefont
  {Caldwell}, \citenamefont {Lindsay}, \citenamefont {Giannini}, \citenamefont
  {Vurgaftman}, \citenamefont {Reinecke}, \citenamefont {Maier},\ and\
  \citenamefont {Glembocki}}]{CaldwellPlasmon2015}%
  \BibitemOpen
  \bibfield  {author} {\bibinfo {author} {\bibfnamefont {J.~D.}\ \bibnamefont
  {Caldwell}}, \bibinfo {author} {\bibfnamefont {L.}~\bibnamefont {Lindsay}},
  \bibinfo {author} {\bibfnamefont {V.}~\bibnamefont {Giannini}}, \bibinfo
  {author} {\bibfnamefont {I.}~\bibnamefont {Vurgaftman}}, \bibinfo {author}
  {\bibfnamefont {T.~L.}\ \bibnamefont {Reinecke}}, \bibinfo {author}
  {\bibfnamefont {S.~A.}\ \bibnamefont {Maier}}, \ and\ \bibinfo {author}
  {\bibfnamefont {O.~J.}\ \bibnamefont {Glembocki}},\ }\href@noop {} {\bibfield
   {journal} {\bibinfo  {journal} {Nanophotonics}\ }\textbf {\bibinfo {volume}
  {4}} (\bibinfo {year} {2015})}\BibitemShut {NoStop}%
\bibitem [{\citenamefont {Pendry}, \citenamefont {Schurig},\ and\ \citenamefont
  {Smith}(2006)}]{Pendry2006}%
  \BibitemOpen
  \bibfield  {author} {\bibinfo {author} {\bibfnamefont {J.~B.}\ \bibnamefont
  {Pendry}}, \bibinfo {author} {\bibfnamefont {D.}~\bibnamefont {Schurig}}, \
  and\ \bibinfo {author} {\bibfnamefont {D.~R.}\ \bibnamefont {Smith}},\
  }\href@noop {} {\bibfield  {journal} {\bibinfo  {journal} {Science}\ }\textbf
  {\bibinfo {volume} {312}},\ \bibinfo {pages} {1780} (\bibinfo {year}
  {2006})}\BibitemShut {NoStop}%
\bibitem [{\citenamefont {Oliveri}, \citenamefont {Werner},\ and\ \citenamefont
  {Massa}(2015)}]{Oliveri2015}%
  \BibitemOpen
  \bibfield  {author} {\bibinfo {author} {\bibfnamefont {G.}~\bibnamefont
  {Oliveri}}, \bibinfo {author} {\bibfnamefont {D.~H.}\ \bibnamefont {Werner}},
  \ and\ \bibinfo {author} {\bibfnamefont {A.}~\bibnamefont {Massa}},\
  }\href@noop {} {\bibfield  {journal} {\bibinfo  {journal} {Proceedings of the
  IEEE}\ }\textbf {\bibinfo {volume} {103}} (\bibinfo {year}
  {2015})}\BibitemShut {NoStop}%
\bibitem [{\citenamefont {Holloway}\ \emph {et~al.}(2012)\citenamefont
  {Holloway}, \citenamefont {Kuester}, \citenamefont {J.~A.~Gordon},
  \citenamefont {Booth},\ and\ \citenamefont {Smith}}]{Holloway2012}%
  \BibitemOpen
  \bibfield  {author} {\bibinfo {author} {\bibfnamefont {C.~L.}\ \bibnamefont
  {Holloway}}, \bibinfo {author} {\bibfnamefont {E.~F.}\ \bibnamefont
  {Kuester}}, \bibinfo {author} {\bibfnamefont {J.~O.}\ \bibnamefont
  {J.~A.~Gordon}}, \bibinfo {author} {\bibfnamefont {J.}~\bibnamefont {Booth}},
  \ and\ \bibinfo {author} {\bibfnamefont {D.~R.}\ \bibnamefont {Smith}},\
  }\href@noop {} {\bibfield  {journal} {\bibinfo  {journal} {IEEE Antennas and
  Propogation Magazine}\ }\textbf {\bibinfo {volume} {54}} (\bibinfo {year}
  {2012})}\BibitemShut {NoStop}%
\bibitem [{\citenamefont {Sarkar}, \citenamefont {Behunin},\ and\ \citenamefont
  {Gibbs}(2019)}]{Sarkar2019}%
  \BibitemOpen
  \bibfield  {author} {\bibinfo {author} {\bibfnamefont {S.}~\bibnamefont
  {Sarkar}}, \bibinfo {author} {\bibfnamefont {R.~O.}\ \bibnamefont {Behunin}},
  \ and\ \bibinfo {author} {\bibfnamefont {J.~G.}\ \bibnamefont {Gibbs}},\
  }\href@noop {} {\bibfield  {journal} {\bibinfo  {journal} {Nano Letters}\
  }\textbf {\bibinfo {volume} {19}} (\bibinfo {year} {2019})}\BibitemShut
  {NoStop}%
\bibitem [{\citenamefont {Liu}\ \emph {et~al.}(2015)\citenamefont {Liu},
  \citenamefont {Jiang}, \citenamefont {Ji}, \citenamefont {Zeng},
  \citenamefont {Zhang}, \citenamefont {Song}, \citenamefont {Xu},\ and\
  \citenamefont {Gan}}]{Liu2015}%
  \BibitemOpen
  \bibfield  {author} {\bibinfo {author} {\bibfnamefont {K.}~\bibnamefont
  {Liu}}, \bibinfo {author} {\bibfnamefont {S.}~\bibnamefont {Jiang}}, \bibinfo
  {author} {\bibfnamefont {D.}~\bibnamefont {Ji}}, \bibinfo {author}
  {\bibfnamefont {X.}~\bibnamefont {Zeng}}, \bibinfo {author} {\bibfnamefont
  {N.}~\bibnamefont {Zhang}}, \bibinfo {author} {\bibfnamefont
  {H.}~\bibnamefont {Song}}, \bibinfo {author} {\bibfnamefont {Y.}~\bibnamefont
  {Xu}}, \ and\ \bibinfo {author} {\bibfnamefont {Q.}~\bibnamefont {Gan}},\
  }\href@noop {} {\bibfield  {journal} {\bibinfo  {journal} {IEEE Photonics
  Technology Letters}\ }\textbf {\bibinfo {volume} {27}} (\bibinfo {year}
  {2015})}\BibitemShut {NoStop}%
\bibitem [{\citenamefont {Baqir}\ and\ \citenamefont
  {Choudhury}(2017)}]{Baqir2017}%
  \BibitemOpen
  \bibfield  {author} {\bibinfo {author} {\bibfnamefont {M.~A.}\ \bibnamefont
  {Baqir}}\ and\ \bibinfo {author} {\bibfnamefont {P.~K.}\ \bibnamefont
  {Choudhury}},\ }\href@noop {} {\bibfield  {journal} {\bibinfo  {journal}
  {IEEE Photonics Technology Letters}\ }\textbf {\bibinfo {volume} {29}}
  (\bibinfo {year} {2017})}\BibitemShut {NoStop}%
\bibitem [{\citenamefont {Dincer}\ \emph {et~al.}(2014)\citenamefont {Dincer},
  \citenamefont {Akgol}, \citenamefont {Karaaslan}, \citenamefont {Unal},\ and\
  \citenamefont {Sabah}}]{Dincer2014}%
  \BibitemOpen
  \bibfield  {author} {\bibinfo {author} {\bibfnamefont {F.}~\bibnamefont
  {Dincer}}, \bibinfo {author} {\bibfnamefont {O.}~\bibnamefont {Akgol}},
  \bibinfo {author} {\bibfnamefont {M.}~\bibnamefont {Karaaslan}}, \bibinfo
  {author} {\bibfnamefont {E.}~\bibnamefont {Unal}}, \ and\ \bibinfo {author}
  {\bibfnamefont {C.}~\bibnamefont {Sabah}},\ }\href@noop {} {\bibfield
  {journal} {\bibinfo  {journal} {Progress in Electromagnetics Research}\
  }\textbf {\bibinfo {volume} {144}} (\bibinfo {year} {2014})}\BibitemShut
  {NoStop}%
\bibitem [{\citenamefont {Rufangura}\ and\ \citenamefont
  {Sabah}(2016)}]{Rufangura2016}%
  \BibitemOpen
  \bibfield  {author} {\bibinfo {author} {\bibfnamefont {P.}~\bibnamefont
  {Rufangura}}\ and\ \bibinfo {author} {\bibfnamefont {C.}~\bibnamefont
  {Sabah}},\ }\href@noop {} {\bibfield  {journal} {\bibinfo  {journal} {Journal
  of Alloys and Compounds}\ }\textbf {\bibinfo {volume} {680}} (\bibinfo {year}
  {2016})}\BibitemShut {NoStop}%
\bibitem [{\citenamefont {Cai}\ \emph {et~al.}(2007)\citenamefont {Cai},
  \citenamefont {Chettiar}, \citenamefont {Kildishev},\ and\ \citenamefont
  {Shalaev}}]{Cai2007}%
  \BibitemOpen
  \bibfield  {author} {\bibinfo {author} {\bibfnamefont {W.}~\bibnamefont
  {Cai}}, \bibinfo {author} {\bibfnamefont {U.~K.}\ \bibnamefont {Chettiar}},
  \bibinfo {author} {\bibfnamefont {A.~V.}\ \bibnamefont {Kildishev}}, \ and\
  \bibinfo {author} {\bibfnamefont {V.~M.}\ \bibnamefont {Shalaev}},\
  }\href@noop {} {\bibfield  {journal} {\bibinfo  {journal} {Nature Photonics}\
  }\textbf {\bibinfo {volume} {1}} (\bibinfo {year} {2007})}\BibitemShut
  {NoStop}%
\bibitem [{\citenamefont {Liu}\ \emph {et~al.}(2008)\citenamefont {Liu},
  \citenamefont {Guo}, \citenamefont {Fu}, \citenamefont {Kaiser},
  \citenamefont {Schweizer},\ and\ \citenamefont {Giessen}}]{Liu2007}%
  \BibitemOpen
  \bibfield  {author} {\bibinfo {author} {\bibfnamefont {N.}~\bibnamefont
  {Liu}}, \bibinfo {author} {\bibfnamefont {H.}~\bibnamefont {Guo}}, \bibinfo
  {author} {\bibfnamefont {L.}~\bibnamefont {Fu}}, \bibinfo {author}
  {\bibfnamefont {S.}~\bibnamefont {Kaiser}}, \bibinfo {author} {\bibfnamefont
  {H.}~\bibnamefont {Schweizer}}, \ and\ \bibinfo {author} {\bibfnamefont
  {H.}~\bibnamefont {Giessen}},\ }\href@noop {} {\bibfield  {journal} {\bibinfo
   {journal} {Nature Materials}\ }\textbf {\bibinfo {volume} {7}} (\bibinfo
  {year} {2008})}\BibitemShut {NoStop}%
\bibitem [{\citenamefont {Carapezza}(2015)}]{Barbosa2015}%
  \BibitemOpen
  \bibinfo {editor} {\bibfnamefont {E.~M.}\ \bibnamefont {Carapezza}},\ ed.,\
  \href@noop {} {\emph {\bibinfo {title} {Proceedings of SPIE}}},\ Vol.\
  \bibinfo {volume} {9456}\ (\bibinfo  {publisher} {SPIE},\ \bibinfo {address}
  {Baltimore, Maryland, United States},\ \bibinfo {year} {2015})\BibitemShut
  {NoStop}%
\bibitem [{\citenamefont {Wang}\ \emph {et~al.}(2015)\citenamefont {Wang},
  \citenamefont {Sivan}, \citenamefont {Mitchell}, \citenamefont
  {Rosengarten},\ and\ \citenamefont {Phelan}}]{Wang2015}%
  \BibitemOpen
  \bibfield  {author} {\bibinfo {author} {\bibfnamefont {H.}~\bibnamefont
  {Wang}}, \bibinfo {author} {\bibfnamefont {V.~P.}\ \bibnamefont {Sivan}},
  \bibinfo {author} {\bibfnamefont {A.}~\bibnamefont {Mitchell}}, \bibinfo
  {author} {\bibfnamefont {G.}~\bibnamefont {Rosengarten}}, \ and\ \bibinfo
  {author} {\bibfnamefont {P.}~\bibnamefont {Phelan}},\ }\href@noop {}
  {\bibfield  {journal} {\bibinfo  {journal} {Solar Energy Materials \& Solar
  Cells}\ }\textbf {\bibinfo {volume} {137}} (\bibinfo {year}
  {2015})}\BibitemShut {NoStop}%
\bibitem [{\citenamefont {Desmet}\ \emph {et~al.}(2006)\citenamefont {Desmet},
  \citenamefont {Paz}, \citenamefont {Corry}, \citenamefont {Eells},
  \citenamefont {Wong-Riley}, \citenamefont {Henry}, \citenamefont {Buchmann},
  \citenamefont {Connelly}, \citenamefont {Dovi}, \citenamefont {Liang},
  \citenamefont {Henshel}, \citenamefont {Yeager}, \citenamefont {Millsap},
  \citenamefont {Lim}, \citenamefont {Gould}, \citenamefont {Das},
  \citenamefont {Jett}, \citenamefont {Hodgson}, \citenamefont {Margolis},\
  and\ \citenamefont {Whelan}}]{Desmet2006}%
  \BibitemOpen
  \bibfield  {author} {\bibinfo {author} {\bibfnamefont {K.~D.}\ \bibnamefont
  {Desmet}}, \bibinfo {author} {\bibfnamefont {D.~A.}\ \bibnamefont {Paz}},
  \bibinfo {author} {\bibfnamefont {J.~J.}\ \bibnamefont {Corry}}, \bibinfo
  {author} {\bibfnamefont {J.~T.}\ \bibnamefont {Eells}}, \bibinfo {author}
  {\bibfnamefont {M.~T.~T.}\ \bibnamefont {Wong-Riley}}, \bibinfo {author}
  {\bibfnamefont {M.~M.}\ \bibnamefont {Henry}}, \bibinfo {author}
  {\bibfnamefont {E.~V.}\ \bibnamefont {Buchmann}}, \bibinfo {author}
  {\bibfnamefont {M.~P.}\ \bibnamefont {Connelly}}, \bibinfo {author}
  {\bibfnamefont {J.~V.}\ \bibnamefont {Dovi}}, \bibinfo {author}
  {\bibfnamefont {H.~L.}\ \bibnamefont {Liang}}, \bibinfo {author}
  {\bibfnamefont {D.~S.}\ \bibnamefont {Henshel}}, \bibinfo {author}
  {\bibfnamefont {R.~L.}\ \bibnamefont {Yeager}}, \bibinfo {author}
  {\bibfnamefont {D.~S.}\ \bibnamefont {Millsap}}, \bibinfo {author}
  {\bibfnamefont {J.}~\bibnamefont {Lim}}, \bibinfo {author} {\bibfnamefont
  {L.~J.}\ \bibnamefont {Gould}}, \bibinfo {author} {\bibfnamefont
  {R.}~\bibnamefont {Das}}, \bibinfo {author} {\bibfnamefont {M.}~\bibnamefont
  {Jett}}, \bibinfo {author} {\bibfnamefont {B.~D.}\ \bibnamefont {Hodgson}},
  \bibinfo {author} {\bibfnamefont {D.}~\bibnamefont {Margolis}}, \ and\
  \bibinfo {author} {\bibfnamefont {H.~T.}\ \bibnamefont {Whelan}},\
  }\href@noop {} {\bibfield  {journal} {\bibinfo  {journal}
  {Photobiomodulation, Photomedicine, and Laser Surgery}\ } (\bibinfo {year}
  {2006})}\BibitemShut {NoStop}%
\bibitem [{\citenamefont {Bellisola}\ and\ \citenamefont
  {Sorio}(2012)}]{Bellisola2012}%
  \BibitemOpen
  \bibfield  {author} {\bibinfo {author} {\bibfnamefont {G.}~\bibnamefont
  {Bellisola}}\ and\ \bibinfo {author} {\bibfnamefont {C.}~\bibnamefont
  {Sorio}},\ }\href@noop {} {\bibfield  {journal} {\bibinfo  {journal}
  {American Journal of Cancer Research}\ }\textbf {\bibinfo {volume} {10}}
  (\bibinfo {year} {2012})}\BibitemShut {NoStop}%
\bibitem [{\citenamefont {Montoya}\ \emph {et~al.}(2017)\citenamefont
  {Montoya}, \citenamefont {Tian}, \citenamefont {Krishna},\ and\ \citenamefont
  {Padilla}}]{Montoya2017}%
  \BibitemOpen
  \bibfield  {author} {\bibinfo {author} {\bibfnamefont {J.~A.}\ \bibnamefont
  {Montoya}}, \bibinfo {author} {\bibfnamefont {Z.}~\bibnamefont {Tian}},
  \bibinfo {author} {\bibfnamefont {S.}~\bibnamefont {Krishna}}, \ and\
  \bibinfo {author} {\bibfnamefont {W.~J.}\ \bibnamefont {Padilla}},\
  }\href@noop {} {\bibfield  {journal} {\bibinfo  {journal} {Optics Express}\
  }\textbf {\bibinfo {volume} {25}} (\bibinfo {year} {2017})}\BibitemShut
  {NoStop}%
\bibitem [{\citenamefont {Ogawa}\ \emph {et~al.}(2012)\citenamefont {Ogawa},
  \citenamefont {Okada}, \citenamefont {Fukushima},\ and\ \citenamefont
  {Kimata}}]{Ogawa2012}%
  \BibitemOpen
  \bibfield  {author} {\bibinfo {author} {\bibfnamefont {S.}~\bibnamefont
  {Ogawa}}, \bibinfo {author} {\bibfnamefont {K.}~\bibnamefont {Okada}},
  \bibinfo {author} {\bibfnamefont {N.}~\bibnamefont {Fukushima}}, \ and\
  \bibinfo {author} {\bibfnamefont {M.}~\bibnamefont {Kimata}},\ }\href@noop {}
  {\bibfield  {journal} {\bibinfo  {journal} {Applied Physics Letters}\
  }\textbf {\bibinfo {volume} {100}} (\bibinfo {year} {2012})}\BibitemShut
  {NoStop}%
\bibitem [{\citenamefont {Xu}\ and\ \citenamefont {Lin}(2020)}]{Xu2020}%
  \BibitemOpen
  \bibfield  {author} {\bibinfo {author} {\bibfnamefont {R.}~\bibnamefont
  {Xu}}\ and\ \bibinfo {author} {\bibfnamefont {Y.}~\bibnamefont {Lin}},\
  }\href@noop {} {\bibfield  {journal} {\bibinfo  {journal} {Nanomaterials}\
  }\textbf {\bibinfo {volume} {10}} (\bibinfo {year} {2020})}\BibitemShut
  {NoStop}%
\bibitem [{\citenamefont {Liu}\ \emph {et~al.}(2010)\citenamefont {Liu},
  \citenamefont {Mesch}, \citenamefont {Weiss}, \citenamefont {Hentschel},\
  and\ \citenamefont {Giessen}}]{Liu2010}%
  \BibitemOpen
  \bibfield  {author} {\bibinfo {author} {\bibfnamefont {N.}~\bibnamefont
  {Liu}}, \bibinfo {author} {\bibfnamefont {M.}~\bibnamefont {Mesch}}, \bibinfo
  {author} {\bibfnamefont {T.}~\bibnamefont {Weiss}}, \bibinfo {author}
  {\bibfnamefont {M.}~\bibnamefont {Hentschel}}, \ and\ \bibinfo {author}
  {\bibfnamefont {H.}~\bibnamefont {Giessen}},\ }\href@noop {} {\bibfield
  {journal} {\bibinfo  {journal} {Nano Letters}\ }\textbf {\bibinfo {volume}
  {10}} (\bibinfo {year} {2010})}\BibitemShut {NoStop}%
\bibitem [{\citenamefont {Xu}, \citenamefont {Xie},\ and\ \citenamefont
  {Ying}(2017)}]{Xu2017THz}%
  \BibitemOpen
  \bibfield  {author} {\bibinfo {author} {\bibfnamefont {W.}~\bibnamefont
  {Xu}}, \bibinfo {author} {\bibfnamefont {L.}~\bibnamefont {Xie}}, \ and\
  \bibinfo {author} {\bibfnamefont {Y.}~\bibnamefont {Ying}},\ }\href@noop {}
  {\bibfield  {journal} {\bibinfo  {journal} {Nanoscale}\ }\textbf {\bibinfo
  {volume} {9}} (\bibinfo {year} {2017})}\BibitemShut {NoStop}%
\bibitem [{\citenamefont {Tonouchi}(2007)}]{Tonouchi2007}%
  \BibitemOpen
  \bibfield  {author} {\bibinfo {author} {\bibfnamefont {M.}~\bibnamefont
  {Tonouchi}},\ }\href@noop {} {\bibfield  {journal} {\bibinfo  {journal}
  {Nature Photonics}\ }\textbf {\bibinfo {volume} {1}},\ \bibinfo {pages} {97}
  (\bibinfo {year} {2007})}\BibitemShut {NoStop}%
\bibitem [{\citenamefont {Son}, \citenamefont {Oh},\ and\ \citenamefont
  {Cheon}(2019)}]{Son2019}%
  \BibitemOpen
  \bibfield  {author} {\bibinfo {author} {\bibfnamefont {J.}~\bibnamefont
  {Son}}, \bibinfo {author} {\bibfnamefont {S.~J.}\ \bibnamefont {Oh}}, \ and\
  \bibinfo {author} {\bibfnamefont {H.}~\bibnamefont {Cheon}},\ }\href@noop {}
  {\bibfield  {journal} {\bibinfo  {journal} {Journal of Applied Physics}\
  }\textbf {\bibinfo {volume} {125}} (\bibinfo {year} {2019})}\BibitemShut
  {NoStop}%
\bibitem [{\citenamefont {Galoda}\ and\ \citenamefont
  {Signh}(2007)}]{Galoda2007}%
  \BibitemOpen
  \bibfield  {author} {\bibinfo {author} {\bibfnamefont {S.}~\bibnamefont
  {Galoda}}\ and\ \bibinfo {author} {\bibfnamefont {G.}~\bibnamefont {Signh}},\
  }\href@noop {} {\bibfield  {journal} {\bibinfo  {journal} {IEEE Potentials}\
  }\textbf {\bibinfo {volume} {26}},\ \bibinfo {pages} {24} (\bibinfo {year}
  {2007})}\BibitemShut {NoStop}%
\bibitem [{\citenamefont {Zhong}(2019)}]{Zhong2019}%
  \BibitemOpen
  \bibfield  {author} {\bibinfo {author} {\bibfnamefont {S.}~\bibnamefont
  {Zhong}},\ }\href@noop {} {\bibfield  {journal} {\bibinfo  {journal}
  {Frontiers of Mechanical Engineering}\ }\textbf {\bibinfo {volume} {14}},\
  \bibinfo {pages} {273} (\bibinfo {year} {2019})}\BibitemShut {NoStop}%
\bibitem [{\citenamefont {Ghanekar}, \citenamefont {Ji},\ and\ \citenamefont
  {Zheng}(2016)}]{Ghanekar2016APL}%
  \BibitemOpen
  \bibfield  {author} {\bibinfo {author} {\bibfnamefont {A.}~\bibnamefont
  {Ghanekar}}, \bibinfo {author} {\bibfnamefont {J.}~\bibnamefont {Ji}}, \ and\
  \bibinfo {author} {\bibfnamefont {Y.}~\bibnamefont {Zheng}},\ }\href@noop {}
  {\bibfield  {journal} {\bibinfo  {journal} {Applied Physics Letters}\
  }\textbf {\bibinfo {volume} {109}} (\bibinfo {year} {2016})}\BibitemShut
  {NoStop}%
\bibitem [{\citenamefont {Ghanekar}\ \emph {et~al.}(2018)\citenamefont
  {Ghanekar}, \citenamefont {Ricci}, \citenamefont {Tian}, \citenamefont
  {Gregory},\ and\ \citenamefont {Zheng}}]{Ghanekar2018APL}%
  \BibitemOpen
  \bibfield  {author} {\bibinfo {author} {\bibfnamefont {A.}~\bibnamefont
  {Ghanekar}}, \bibinfo {author} {\bibfnamefont {M.}~\bibnamefont {Ricci}},
  \bibinfo {author} {\bibfnamefont {Y.}~\bibnamefont {Tian}}, \bibinfo {author}
  {\bibfnamefont {O.}~\bibnamefont {Gregory}}, \ and\ \bibinfo {author}
  {\bibfnamefont {Y.}~\bibnamefont {Zheng}},\ }\href@noop {} {\bibfield
  {journal} {\bibinfo  {journal} {Applied Physics Letters}\ }\textbf {\bibinfo
  {volume} {112}} (\bibinfo {year} {2018})}\BibitemShut {NoStop}%
\bibitem [{\citenamefont {Turpin}\ \emph {et~al.}(2014)\citenamefont {Turpin},
  \citenamefont {Bossard}, \citenamefont {Morgan}, \citenamefont {Werner},\
  and\ \citenamefont {Werner}}]{Turpin2014}%
  \BibitemOpen
  \bibfield  {author} {\bibinfo {author} {\bibfnamefont {J.~P.}\ \bibnamefont
  {Turpin}}, \bibinfo {author} {\bibfnamefont {J.~A.}\ \bibnamefont {Bossard}},
  \bibinfo {author} {\bibfnamefont {K.~L.}\ \bibnamefont {Morgan}}, \bibinfo
  {author} {\bibfnamefont {D.~H.}\ \bibnamefont {Werner}}, \ and\ \bibinfo
  {author} {\bibfnamefont {P.~L.}\ \bibnamefont {Werner}},\ }\href@noop {}
  {\bibfield  {journal} {\bibinfo  {journal} {International Journal of Antennas
  and Propagation}\ }\textbf {\bibinfo {volume} {2014}} (\bibinfo {year}
  {2014})}\BibitemShut {NoStop}%
\bibitem [{\citenamefont {Cao}\ \emph {et~al.}(2018)\citenamefont {Cao},
  \citenamefont {Zhang}, \citenamefont {Dong}, \citenamefont {Lu},
  \citenamefont {Zhou}, \citenamefont {Zhuang}, \citenamefont {Deng},
  \citenamefont {Cheng}, \citenamefont {Li},\ and\ \citenamefont
  {Simpson}}]{Cao2018}%
  \BibitemOpen
  \bibfield  {author} {\bibinfo {author} {\bibfnamefont {T.}~\bibnamefont
  {Cao}}, \bibinfo {author} {\bibfnamefont {X.}~\bibnamefont {Zhang}}, \bibinfo
  {author} {\bibfnamefont {W.}~\bibnamefont {Dong}}, \bibinfo {author}
  {\bibfnamefont {L.}~\bibnamefont {Lu}}, \bibinfo {author} {\bibfnamefont
  {X.}~\bibnamefont {Zhou}}, \bibinfo {author} {\bibfnamefont {X.}~\bibnamefont
  {Zhuang}}, \bibinfo {author} {\bibfnamefont {J.}~\bibnamefont {Deng}},
  \bibinfo {author} {\bibfnamefont {X.}~\bibnamefont {Cheng}}, \bibinfo
  {author} {\bibfnamefont {G.}~\bibnamefont {Li}}, \ and\ \bibinfo {author}
  {\bibfnamefont {R.~E.}\ \bibnamefont {Simpson}},\ }\href@noop {} {\bibfield
  {journal} {\bibinfo  {journal} {Advanced Optical Materials}\ }\textbf
  {\bibinfo {volume} {6}} (\bibinfo {year} {2018})}\BibitemShut {NoStop}%
\bibitem [{\citenamefont {Qu}\ \emph {et~al.}(2017)\citenamefont {Qu},
  \citenamefont {Li}, \citenamefont {Du}, \citenamefont {Cai}, \citenamefont
  {Lu},\ and\ \citenamefont {Qiu}}]{Qu2017}%
  \BibitemOpen
  \bibfield  {author} {\bibinfo {author} {\bibfnamefont {Y.}~\bibnamefont
  {Qu}}, \bibinfo {author} {\bibfnamefont {Q.}~\bibnamefont {Li}}, \bibinfo
  {author} {\bibfnamefont {K.}~\bibnamefont {Du}}, \bibinfo {author}
  {\bibfnamefont {L.}~\bibnamefont {Cai}}, \bibinfo {author} {\bibfnamefont
  {J.}~\bibnamefont {Lu}}, \ and\ \bibinfo {author} {\bibfnamefont
  {M.}~\bibnamefont {Qiu}},\ }\href@noop {} {\bibfield  {journal} {\bibinfo
  {journal} {Laser and Photonics Reviews}\ }\textbf {\bibinfo {volume} {11}}
  (\bibinfo {year} {2017})}\BibitemShut {NoStop}%
\bibitem [{\citenamefont {Nguyen}\ \emph {et~al.}(2020)\citenamefont {Nguyen},
  \citenamefont {Biu}, \citenamefont {Nguyen}, \citenamefont {Vu},\ and\
  \citenamefont {Bui}}]{Nguyen2020}%
  \BibitemOpen
  \bibfield  {author} {\bibinfo {author} {\bibfnamefont {T.~H.}\ \bibnamefont
  {Nguyen}}, \bibinfo {author} {\bibfnamefont {S.~T.}\ \bibnamefont {Biu}},
  \bibinfo {author} {\bibfnamefont {X.~C.}\ \bibnamefont {Nguyen}}, \bibinfo
  {author} {\bibfnamefont {D.~L.}\ \bibnamefont {Vu}}, \ and\ \bibinfo {author}
  {\bibfnamefont {X.~K.}\ \bibnamefont {Bui}},\ }\href@noop {} {\bibfield
  {journal} {\bibinfo  {journal} {RSC Advances}\ } (\bibinfo {year}
  {2020})}\BibitemShut {NoStop}%
\bibitem [{\citenamefont {Xu}\ \emph {et~al.}(2018)\citenamefont {Xu},
  \citenamefont {Luo}, \citenamefont {Sha}, \citenamefont {Zhong},
  \citenamefont {Xu}, \citenamefont {Tong},\ and\ \citenamefont
  {Lin}}]{Xu2018-deformation}%
  \BibitemOpen
  \bibfield  {author} {\bibinfo {author} {\bibfnamefont {R.}~\bibnamefont
  {Xu}}, \bibinfo {author} {\bibfnamefont {J.}~\bibnamefont {Luo}}, \bibinfo
  {author} {\bibfnamefont {J.}~\bibnamefont {Sha}}, \bibinfo {author}
  {\bibfnamefont {J.}~\bibnamefont {Zhong}}, \bibinfo {author} {\bibfnamefont
  {Z.}~\bibnamefont {Xu}}, \bibinfo {author} {\bibfnamefont {Y.}~\bibnamefont
  {Tong}}, \ and\ \bibinfo {author} {\bibfnamefont {Y.}~\bibnamefont {Lin}},\
  }\href@noop {} {\bibfield  {journal} {\bibinfo  {journal} {Applied Physics
  Letters}\ }\textbf {\bibinfo {volume} {113}} (\bibinfo {year}
  {2018})}\BibitemShut {NoStop}%
\bibitem [{\citenamefont {Wang}\ \emph {et~al.}(2020)\citenamefont {Wang},
  \citenamefont {Lv}, \citenamefont {Becton}, \citenamefont {Hong},
  \citenamefont {Zhang},\ and\ \citenamefont {Chen}}]{Wang2020}%
  \BibitemOpen
  \bibfield  {author} {\bibinfo {author} {\bibfnamefont {Z.}~\bibnamefont
  {Wang}}, \bibinfo {author} {\bibfnamefont {P.}~\bibnamefont {Lv}}, \bibinfo
  {author} {\bibfnamefont {M.}~\bibnamefont {Becton}}, \bibinfo {author}
  {\bibfnamefont {J.}~\bibnamefont {Hong}}, \bibinfo {author} {\bibfnamefont
  {L.}~\bibnamefont {Zhang}}, \ and\ \bibinfo {author} {\bibfnamefont
  {X.}~\bibnamefont {Chen}},\ }\href@noop {} {\bibfield  {journal} {\bibinfo
  {journal} {Langmuir}\ }\textbf {\bibinfo {volume} {36}} (\bibinfo {year}
  {2020})}\BibitemShut {NoStop}%
\bibitem [{\citenamefont {Su}\ \emph {et~al.}(2020)\citenamefont {Su},
  \citenamefont {Feng}, \citenamefont {Zeng},\ and\ \citenamefont
  {Yu}}]{Su2020}%
  \BibitemOpen
  \bibfield  {author} {\bibinfo {author} {\bibfnamefont {X.}~\bibnamefont
  {Su}}, \bibinfo {author} {\bibfnamefont {C.}~\bibnamefont {Feng}}, \bibinfo
  {author} {\bibfnamefont {Y.}~\bibnamefont {Zeng}}, \ and\ \bibinfo {author}
  {\bibfnamefont {H.}~\bibnamefont {Yu}},\ }\href@noop {} {\bibfield  {journal}
  {\bibinfo  {journal} {Optics Communications}\ }\textbf {\bibinfo {volume}
  {459}} (\bibinfo {year} {2020})}\BibitemShut {NoStop}%
\bibitem [{\citenamefont {Chen}, \citenamefont {Cheng},\ and\ \citenamefont
  {Luo}(2020)}]{Chen2020}%
  \BibitemOpen
  \bibfield  {author} {\bibinfo {author} {\bibfnamefont {F.}~\bibnamefont
  {Chen}}, \bibinfo {author} {\bibfnamefont {Y.}~\bibnamefont {Cheng}}, \ and\
  \bibinfo {author} {\bibfnamefont {H.}~\bibnamefont {Luo}},\ }\href@noop {}
  {\bibfield  {journal} {\bibinfo  {journal} {Materials}\ }\textbf {\bibinfo
  {volume} {13}} (\bibinfo {year} {2020})}\BibitemShut {NoStop}%
\bibitem [{\citenamefont {Qi}\ \emph {et~al.}(2020)\citenamefont {Qi},
  \citenamefont {Zhang}, \citenamefont {Liu}, \citenamefont {Zhang},
  \citenamefont {Zhang}, \citenamefont {Wang}, \citenamefont {Deng},
  \citenamefont {Wang},\ and\ \citenamefont {Yu}}]{Qi2020}%
  \BibitemOpen
  \bibfield  {author} {\bibinfo {author} {\bibfnamefont {Y.}~\bibnamefont
  {Qi}}, \bibinfo {author} {\bibfnamefont {Y.}~\bibnamefont {Zhang}}, \bibinfo
  {author} {\bibfnamefont {C.}~\bibnamefont {Liu}}, \bibinfo {author}
  {\bibfnamefont {T.}~\bibnamefont {Zhang}}, \bibinfo {author} {\bibfnamefont
  {B.}~\bibnamefont {Zhang}}, \bibinfo {author} {\bibfnamefont
  {L.}~\bibnamefont {Wang}}, \bibinfo {author} {\bibfnamefont {X.}~\bibnamefont
  {Deng}}, \bibinfo {author} {\bibfnamefont {X.}~\bibnamefont {Wang}}, \ and\
  \bibinfo {author} {\bibfnamefont {Y.}~\bibnamefont {Yu}},\ }\href@noop {}
  {\bibfield  {journal} {\bibinfo  {journal} {Nanomaterials}\ }\textbf
  {\bibinfo {volume} {10}} (\bibinfo {year} {2020})}\BibitemShut {NoStop}%
\bibitem [{\citenamefont {Moncada-Villa}\ \emph {et~al.}(2015)\citenamefont
  {Moncada-Villa}, \citenamefont {Fernanandez-Hurtado}, \citenamefont
  {Garcia-Vidal}, \citenamefont {Garcia-Martin},\ and\ \citenamefont
  {Cuevas}}]{Moncada2015}%
  \BibitemOpen
  \bibfield  {author} {\bibinfo {author} {\bibfnamefont {E.}~\bibnamefont
  {Moncada-Villa}}, \bibinfo {author} {\bibfnamefont {V.}~\bibnamefont
  {Fernanandez-Hurtado}}, \bibinfo {author} {\bibfnamefont {F.~J.}\
  \bibnamefont {Garcia-Vidal}}, \bibinfo {author} {\bibfnamefont
  {A.}~\bibnamefont {Garcia-Martin}}, \ and\ \bibinfo {author} {\bibfnamefont
  {J.~C.}\ \bibnamefont {Cuevas}},\ }\href@noop {} {\bibfield  {journal}
  {\bibinfo  {journal} {Physical Review B}\ }\textbf {\bibinfo {volume} {92}}
  (\bibinfo {year} {2015})}\BibitemShut {NoStop}%
\bibitem [{\citenamefont {Hu}, \citenamefont {Zhang},\ and\ \citenamefont
  {Wang}(2015)}]{Hu2015}%
  \BibitemOpen
  \bibfield  {author} {\bibinfo {author} {\bibfnamefont {B.}~\bibnamefont
  {Hu}}, \bibinfo {author} {\bibfnamefont {Y.}~\bibnamefont {Zhang}}, \ and\
  \bibinfo {author} {\bibfnamefont {Q.~J.}\ \bibnamefont {Wang}},\ }\href@noop
  {} {\bibfield  {journal} {\bibinfo  {journal} {Nanomaterials}\ }\textbf
  {\bibinfo {volume} {4}} (\bibinfo {year} {2015})}\BibitemShut {NoStop}%
\bibitem [{\citenamefont {Lee}, \citenamefont {Choi},\ and\ \citenamefont
  {Jeong}(2020)}]{Lee2020}%
  \BibitemOpen
  \bibfield  {author} {\bibinfo {author} {\bibfnamefont {C.}~\bibnamefont
  {Lee}}, \bibinfo {author} {\bibfnamefont {H.~J.}\ \bibnamefont {Choi}}, \
  and\ \bibinfo {author} {\bibfnamefont {H.}~\bibnamefont {Jeong}},\
  }\href@noop {} {\bibfield  {journal} {\bibinfo  {journal} {Nano Convergence}\
  }\textbf {\bibinfo {volume} {7}} (\bibinfo {year} {2020})}\BibitemShut
  {NoStop}%
\bibitem [{\citenamefont {Dicken}\ \emph {et~al.}(2009)\citenamefont {Dicken},
  \citenamefont {Aydin}, \citenamefont {Pryce}, \citenamefont {Sweatlock},
  \citenamefont {Boyd}, \citenamefont {S.~Walavalkar},\ and\ \citenamefont
  {Atwater}}]{Dicken2009}%
  \BibitemOpen
  \bibfield  {author} {\bibinfo {author} {\bibfnamefont {M.~J.}\ \bibnamefont
  {Dicken}}, \bibinfo {author} {\bibfnamefont {K.}~\bibnamefont {Aydin}},
  \bibinfo {author} {\bibfnamefont {I.~M.}\ \bibnamefont {Pryce}}, \bibinfo
  {author} {\bibfnamefont {L.~A.}\ \bibnamefont {Sweatlock}}, \bibinfo {author}
  {\bibfnamefont {E.~M.}\ \bibnamefont {Boyd}}, \bibinfo {author}
  {\bibfnamefont {J.~M.}\ \bibnamefont {S.~Walavalkar}}, \ and\ \bibinfo
  {author} {\bibfnamefont {H.~A.}\ \bibnamefont {Atwater}},\ }\href@noop {}
  {\bibfield  {journal} {\bibinfo  {journal} {Optics Express}\ }\textbf
  {\bibinfo {volume} {17}} (\bibinfo {year} {2009})}\BibitemShut {NoStop}%
\bibitem [{\citenamefont {Li}\ \emph {et~al.}(2019)\citenamefont {Li},
  \citenamefont {Shen}, \citenamefont {Cao}, \citenamefont {Zhang},\ and\
  \citenamefont {Meng}}]{Li2019APL}%
  \BibitemOpen
  \bibfield  {author} {\bibinfo {author} {\bibfnamefont {Y.}~\bibnamefont
  {Li}}, \bibinfo {author} {\bibfnamefont {Y.}~\bibnamefont {Shen}}, \bibinfo
  {author} {\bibfnamefont {S.}~\bibnamefont {Cao}}, \bibinfo {author}
  {\bibfnamefont {X.}~\bibnamefont {Zhang}}, \ and\ \bibinfo {author}
  {\bibfnamefont {Y.}~\bibnamefont {Meng}},\ }\href@noop {} {\bibfield
  {journal} {\bibinfo  {journal} {Applied Physics Letters}\ }\textbf {\bibinfo
  {volume} {114}} (\bibinfo {year} {2019})}\BibitemShut {NoStop}%
\bibitem [{\citenamefont {Jia}\ \emph {et~al.}(2016)\citenamefont {Jia},
  \citenamefont {Wang}, \citenamefont {Yuan}, \citenamefont {Meng},\ and\
  \citenamefont {Zhou}}]{Jia2016APL}%
  \BibitemOpen
  \bibfield  {author} {\bibinfo {author} {\bibfnamefont {X.}~\bibnamefont
  {Jia}}, \bibinfo {author} {\bibfnamefont {X.}~\bibnamefont {Wang}}, \bibinfo
  {author} {\bibfnamefont {C.}~\bibnamefont {Yuan}}, \bibinfo {author}
  {\bibfnamefont {Q.}~\bibnamefont {Meng}}, \ and\ \bibinfo {author}
  {\bibfnamefont {Z.}~\bibnamefont {Zhou}},\ }\href@noop {} {\bibfield
  {journal} {\bibinfo  {journal} {Applied Physics Letters}\ }\textbf {\bibinfo
  {volume} {120}} (\bibinfo {year} {2016})}\BibitemShut {NoStop}%
\bibitem [{\citenamefont {Driscoll}\ \emph {et~al.}(2008)\citenamefont
  {Driscoll}, \citenamefont {Palit}, \citenamefont {Qazilbash}, \citenamefont
  {Brehm}, \citenamefont {Keilmann}, \citenamefont {Chae}, \citenamefont {Yun},
  \citenamefont {Kim}, \citenamefont {Cho}, \citenamefont {Jokerst},
  \citenamefont {Smith},\ and\ \citenamefont {Basov}}]{Driscoll2008APL}%
  \BibitemOpen
  \bibfield  {author} {\bibinfo {author} {\bibfnamefont {T.}~\bibnamefont
  {Driscoll}}, \bibinfo {author} {\bibfnamefont {S.}~\bibnamefont {Palit}},
  \bibinfo {author} {\bibfnamefont {M.~M.}\ \bibnamefont {Qazilbash}}, \bibinfo
  {author} {\bibfnamefont {M.}~\bibnamefont {Brehm}}, \bibinfo {author}
  {\bibfnamefont {F.}~\bibnamefont {Keilmann}}, \bibinfo {author}
  {\bibfnamefont {B.}~\bibnamefont {Chae}}, \bibinfo {author} {\bibfnamefont
  {S.}~\bibnamefont {Yun}}, \bibinfo {author} {\bibfnamefont {H.}~\bibnamefont
  {Kim}}, \bibinfo {author} {\bibfnamefont {S.~Y.}\ \bibnamefont {Cho}},
  \bibinfo {author} {\bibfnamefont {N.~M.}\ \bibnamefont {Jokerst}}, \bibinfo
  {author} {\bibfnamefont {D.~R.}\ \bibnamefont {Smith}}, \ and\ \bibinfo
  {author} {\bibfnamefont {D.~N.}\ \bibnamefont {Basov}},\ }\href@noop {}
  {\bibfield  {journal} {\bibinfo  {journal} {Applied Physics Letters}\
  }\textbf {\bibinfo {volume} {93}} (\bibinfo {year} {2008})}\BibitemShut
  {NoStop}%
\bibitem [{\citenamefont {Gutruf}\ \emph {et~al.}(2016)\citenamefont {Gutruf},
  \citenamefont {Zou}, \citenamefont {Withayachumnankul}, \citenamefont
  {Bhaskaran}, \citenamefont {Sriram},\ and\ \citenamefont
  {Fumeaux}}]{Gutruf2016}%
  \BibitemOpen
  \bibfield  {author} {\bibinfo {author} {\bibfnamefont {P.}~\bibnamefont
  {Gutruf}}, \bibinfo {author} {\bibfnamefont {C.}~\bibnamefont {Zou}},
  \bibinfo {author} {\bibfnamefont {W.}~\bibnamefont {Withayachumnankul}},
  \bibinfo {author} {\bibfnamefont {M.}~\bibnamefont {Bhaskaran}}, \bibinfo
  {author} {\bibfnamefont {S.}~\bibnamefont {Sriram}}, \ and\ \bibinfo {author}
  {\bibfnamefont {C.}~\bibnamefont {Fumeaux}},\ }\href@noop {} {\bibfield
  {journal} {\bibinfo  {journal} {ACS Nano}\ }\textbf {\bibinfo {volume}
  {10}},\ \bibinfo {pages} {133} (\bibinfo {year} {2016})}\BibitemShut
  {NoStop}%
\bibitem [{\citenamefont {Liu}\ \emph {et~al.}(2020)\citenamefont {Liu},
  \citenamefont {Tian}, \citenamefont {Chen}, \citenamefont {Ghanekar},
  \citenamefont {Antezza},\ and\ \citenamefont {Zheng}}]{Liu2020Nature}%
  \BibitemOpen
  \bibfield  {author} {\bibinfo {author} {\bibfnamefont {X.}~\bibnamefont
  {Liu}}, \bibinfo {author} {\bibfnamefont {Y.}~\bibnamefont {Tian}}, \bibinfo
  {author} {\bibfnamefont {F.}~\bibnamefont {Chen}}, \bibinfo {author}
  {\bibfnamefont {A.}~\bibnamefont {Ghanekar}}, \bibinfo {author}
  {\bibfnamefont {M.}~\bibnamefont {Antezza}}, \ and\ \bibinfo {author}
  {\bibfnamefont {Y.}~\bibnamefont {Zheng}},\ }\href@noop {} {\bibfield
  {journal} {\bibinfo  {journal} {Communications Materials}\ }\textbf {\bibinfo
  {volume} {1}} (\bibinfo {year} {2020})}\BibitemShut {NoStop}%
\bibitem [{\citenamefont {Sharma}\ \emph {et~al.}(2020)\citenamefont {Sharma},
  \citenamefont {Lakhtakia}, \citenamefont {Bhattacharyya},\ and\ \citenamefont
  {Jain}}]{Sharma2020}%
  \BibitemOpen
  \bibfield  {author} {\bibinfo {author} {\bibfnamefont {G.}~\bibnamefont
  {Sharma}}, \bibinfo {author} {\bibfnamefont {A.}~\bibnamefont {Lakhtakia}},
  \bibinfo {author} {\bibfnamefont {S.}~\bibnamefont {Bhattacharyya}}, \ and\
  \bibinfo {author} {\bibfnamefont {P.~K.}\ \bibnamefont {Jain}},\ }\href@noop
  {} {\bibfield  {journal} {\bibinfo  {journal} {Applied Optics}\ }\textbf
  {\bibinfo {volume} {59}} (\bibinfo {year} {2020})}\BibitemShut {NoStop}%
\bibitem [{\citenamefont {Li}\ \emph {et~al.}(2013)\citenamefont {Li},
  \citenamefont {Ma}, \citenamefont {Zhang}, \citenamefont {Wang},
  \citenamefont {Hu}, \citenamefont {Xu},\ and\ \citenamefont
  {Song}}]{Li2013Thz}%
  \BibitemOpen
  \bibfield  {author} {\bibinfo {author} {\bibfnamefont {K.}~\bibnamefont
  {Li}}, \bibinfo {author} {\bibfnamefont {X.}~\bibnamefont {Ma}}, \bibinfo
  {author} {\bibfnamefont {Z.}~\bibnamefont {Zhang}}, \bibinfo {author}
  {\bibfnamefont {L.}~\bibnamefont {Wang}}, \bibinfo {author} {\bibfnamefont
  {H.}~\bibnamefont {Hu}}, \bibinfo {author} {\bibfnamefont {Y.}~\bibnamefont
  {Xu}}, \ and\ \bibinfo {author} {\bibfnamefont {G.}~\bibnamefont {Song}},\
  }\href@noop {} {\bibfield  {journal} {\bibinfo  {journal} {AIP Advances}\
  }\textbf {\bibinfo {volume} {3}} (\bibinfo {year} {2013})}\BibitemShut
  {NoStop}%
\bibitem [{\citenamefont {Doyeux}\ \emph {et~al.}(2017)\citenamefont {Doyeux},
  \citenamefont {Gangaraj}, \citenamefont {Hanson},\ and\ \citenamefont
  {Antezza}}]{Mauro1}%
  \BibitemOpen
  \bibfield  {author} {\bibinfo {author} {\bibfnamefont {P.}~\bibnamefont
  {Doyeux}}, \bibinfo {author} {\bibfnamefont {S.~A.~H.}\ \bibnamefont
  {Gangaraj}}, \bibinfo {author} {\bibfnamefont {G.~W.}\ \bibnamefont
  {Hanson}}, \ and\ \bibinfo {author} {\bibfnamefont {M.}~\bibnamefont
  {Antezza}},\ }\href@noop {} {\bibfield  {journal} {\bibinfo  {journal}
  {Physical Review Letters}\ }\textbf {\bibinfo {volume} {119}} (\bibinfo
  {year} {2017})}\BibitemShut {NoStop}%
\bibitem [{\citenamefont {Gangaraj}, \citenamefont {Hanson},\ and\
  \citenamefont {Antezza}(2017)}]{Mauro2}%
  \BibitemOpen
  \bibfield  {author} {\bibinfo {author} {\bibfnamefont {S.~A.~H.}\
  \bibnamefont {Gangaraj}}, \bibinfo {author} {\bibfnamefont {G.~W.}\
  \bibnamefont {Hanson}}, \ and\ \bibinfo {author} {\bibfnamefont
  {M.}~\bibnamefont {Antezza}},\ }\href@noop {} {\bibfield  {journal} {\bibinfo
   {journal} {Physical Review A}\ }\textbf {\bibinfo {volume} {95}} (\bibinfo
  {year} {2017})}\BibitemShut {NoStop}%
\bibitem [{\citenamefont {Ghanekar}, \citenamefont {Lin},\ and\ \citenamefont
  {Zheng}(2016)}]{Ghanekar2016}%
  \BibitemOpen
  \bibfield  {author} {\bibinfo {author} {\bibfnamefont {A.}~\bibnamefont
  {Ghanekar}}, \bibinfo {author} {\bibfnamefont {L.}~\bibnamefont {Lin}}, \
  and\ \bibinfo {author} {\bibfnamefont {Y.}~\bibnamefont {Zheng}},\
  }\href@noop {} {\bibfield  {journal} {\bibinfo  {journal} {Optics Express}\
  }\textbf {\bibinfo {volume} {24}} (\bibinfo {year} {2016})}\BibitemShut
  {NoStop}%
\bibitem [{\citenamefont {Tian}\ \emph {et~al.}(2020)\citenamefont {Tian},
  \citenamefont {Liu}, \citenamefont {Ghanekar}, \citenamefont {Chen},
  \citenamefont {Caratenuto},\ and\ \citenamefont {Zheng}}]{Tian2020}%
  \BibitemOpen
  \bibfield  {author} {\bibinfo {author} {\bibfnamefont {Y.}~\bibnamefont
  {Tian}}, \bibinfo {author} {\bibfnamefont {X.}~\bibnamefont {Liu}}, \bibinfo
  {author} {\bibfnamefont {A.}~\bibnamefont {Ghanekar}}, \bibinfo {author}
  {\bibfnamefont {F.}~\bibnamefont {Chen}}, \bibinfo {author} {\bibfnamefont
  {A.}~\bibnamefont {Caratenuto}}, \ and\ \bibinfo {author} {\bibfnamefont
  {Y.}~\bibnamefont {Zheng}},\ }\href@noop {} {\bibfield  {journal} {\bibinfo
  {journal} {Scientific Reports}\ }\textbf {\bibinfo {volume} {10}} (\bibinfo
  {year} {2020})}\BibitemShut {NoStop}%
\bibitem [{\citenamefont {Zhang}, \citenamefont {Wu},\ and\ \citenamefont
  {Fu}(2020)}]{Zhang_Kirchoff}%
  \BibitemOpen
  \bibfield  {author} {\bibinfo {author} {\bibfnamefont {Z.~M.}\ \bibnamefont
  {Zhang}}, \bibinfo {author} {\bibfnamefont {X.}~\bibnamefont {Wu}}, \ and\
  \bibinfo {author} {\bibfnamefont {C.}~\bibnamefont {Fu}},\ }\href@noop {}
  {\bibfield  {journal} {\bibinfo  {journal} {Journal of Quantitative
  Spectroscopy and Radiative Transfer}\ }\textbf {\bibinfo {volume} {245}}
  (\bibinfo {year} {2020})}\BibitemShut {NoStop}%
\bibitem [{\citenamefont {Zhu}\ and\ \citenamefont {Fan}(2014)}]{Zhu_Kirchoff}%
  \BibitemOpen
  \bibfield  {author} {\bibinfo {author} {\bibfnamefont {L.}~\bibnamefont
  {Zhu}}\ and\ \bibinfo {author} {\bibfnamefont {S.}~\bibnamefont {Fan}},\
  }\href@noop {} {\bibfield  {journal} {\bibinfo  {journal} {Physical Review
  B}\ }\textbf {\bibinfo {volume} {90}} (\bibinfo {year} {2014})}\BibitemShut
  {NoStop}%
\bibitem [{\citenamefont {Zhao}\ \emph {et~al.}(2019)\citenamefont {Zhao},
  \citenamefont {Shi}, \citenamefont {Wang}, \citenamefont {Zhao},
  \citenamefont {Zhao},\ and\ \citenamefont {Fan}}]{Zhao_Kirchoff}%
  \BibitemOpen
  \bibfield  {author} {\bibinfo {author} {\bibfnamefont {B.}~\bibnamefont
  {Zhao}}, \bibinfo {author} {\bibfnamefont {Y.}~\bibnamefont {Shi}}, \bibinfo
  {author} {\bibfnamefont {J.}~\bibnamefont {Wang}}, \bibinfo {author}
  {\bibfnamefont {Z.}~\bibnamefont {Zhao}}, \bibinfo {author} {\bibfnamefont
  {N.}~\bibnamefont {Zhao}}, \ and\ \bibinfo {author} {\bibfnamefont
  {S.}~\bibnamefont {Fan}},\ }\href@noop {} {\bibfield  {journal} {\bibinfo
  {journal} {Optics Letters}\ }\textbf {\bibinfo {volume} {44}} (\bibinfo
  {year} {2019})}\BibitemShut {NoStop}%
\bibitem [{\citenamefont {Ordal}\ \emph {et~al.}(1988)\citenamefont {Ordal},
  \citenamefont {Bell}, \citenamefont {Alexander}, \citenamefont {Newquist},\
  and\ \citenamefont {Querry}}]{Ordal1988}%
  \BibitemOpen
  \bibfield  {author} {\bibinfo {author} {\bibfnamefont {M.~A.}\ \bibnamefont
  {Ordal}}, \bibinfo {author} {\bibfnamefont {R.~J.}\ \bibnamefont {Bell}},
  \bibinfo {author} {\bibfnamefont {R.~W.}\ \bibnamefont {Alexander}}, \bibinfo
  {author} {\bibfnamefont {L.~A.}\ \bibnamefont {Newquist}}, \ and\ \bibinfo
  {author} {\bibfnamefont {M.~R.}\ \bibnamefont {Querry}},\ }\href@noop {}
  {\bibfield  {journal} {\bibinfo  {journal} {Applied Optics}\ }\textbf
  {\bibinfo {volume} {27}},\ \bibinfo {pages} {1203} (\bibinfo {year}
  {1988})}\BibitemShut {NoStop}%
\bibitem [{\citenamefont {Palik}\ \emph {et~al.}(1976)\citenamefont {Palik},
  \citenamefont {Kaplan}, \citenamefont {Gammon}, \citenamefont {Kaplan},
  \citenamefont {Wallis},\ and\ \citenamefont {Quinn}}]{Palik1976}%
  \BibitemOpen
  \bibfield  {author} {\bibinfo {author} {\bibfnamefont {E.~D.}\ \bibnamefont
  {Palik}}, \bibinfo {author} {\bibfnamefont {R.}~\bibnamefont {Kaplan}},
  \bibinfo {author} {\bibfnamefont {R.~W.}\ \bibnamefont {Gammon}}, \bibinfo
  {author} {\bibfnamefont {H.}~\bibnamefont {Kaplan}}, \bibinfo {author}
  {\bibfnamefont {R.~F.}\ \bibnamefont {Wallis}}, \ and\ \bibinfo {author}
  {\bibfnamefont {J.~J.}\ \bibnamefont {Quinn}},\ }\href@noop {} {\bibfield
  {journal} {\bibinfo  {journal} {Physical Review B}\ }\textbf {\bibinfo
  {volume} {13}} (\bibinfo {year} {1976})}\BibitemShut {NoStop}%
\bibitem [{\citenamefont {Ayala}(2007)}]{Ayala2007}%
  \BibitemOpen
  \bibfield  {author} {\bibinfo {author} {\bibfnamefont {A.~P.}\ \bibnamefont
  {Ayala}},\ }\href@noop {} {\bibfield  {journal} {\bibinfo  {journal}
  {Vibrational Spectroscopy}\ }\textbf {\bibinfo {volume} {45}},\ \bibinfo
  {pages} {112} (\bibinfo {year} {2007})}\BibitemShut {NoStop}%
\bibitem [{Bet(1996)}]{Betz1996}%
  \BibitemOpen
  \href@noop {} {\emph {\bibinfo {title} {A Practical Schottky Mixer for 5 THz
  (Part II)}}}\ (\bibinfo  {publisher} {Seventh International Symposium on
  Space Terahertz Technology},\ \bibinfo {address} {Charlottesville},\ \bibinfo
  {year} {1996})\BibitemShut {NoStop}%
\bibitem [{\citenamefont {Mollaoglu}, \citenamefont {Ozyurt},\ and\
  \citenamefont {Severcan}(2019)}]{Mollaoglu2019}%
  \BibitemOpen
  \bibfield  {author} {\bibinfo {author} {\bibfnamefont {A.~D.}\ \bibnamefont
  {Mollaoglu}}, \bibinfo {author} {\bibfnamefont {I.}~\bibnamefont {Ozyurt}}, \
  and\ \bibinfo {author} {\bibfnamefont {F.}~\bibnamefont {Severcan}},\
  }\enquote {\bibinfo {title} {Infrared spectroscopy: Principles, advances and
  applications},}\ \ (\bibinfo  {publisher} {IntechOpen},\ \bibinfo {year}
  {2019})\ Chap.~\bibinfo {chapter} {5}\BibitemShut {NoStop}%
\end{thebibliography}

\providecommand{\noopsort}[1]{}\providecommand{\singleletter}[1]{#1}%

\end{document}